\documentclass[10pt,twocolumn,letterpaper]{article}

\usepackage{iccv}
\usepackage{times}
\usepackage{epsfig}
\usepackage{graphicx}
\usepackage{amsmath}
\usepackage{amssymb}

\usepackage{pdfpages}
\usepackage{tabularx}
\usepackage{booktabs} 
\usepackage{amssymb}
\usepackage{xcolor}
\usepackage{pifont}     

\usepackage{caption}
\usepackage{graphicx}
\usepackage{subcaption}
\usepackage[pagebackref=true,breaklinks=true,letterpaper=true,colorlinks,bookmarks=false]{hyperref}

\usepackage[accsupp]{axessibility}  

\iccvfinalcopy 

\ificcvfinal\fi

\begin{document}

\definecolor{maroon}{rgb}{0.80,0.,0.} 
\definecolor{myred}{rgb}{0.8,0,0}
\definecolor{mygreen}{rgb}{0,0.8,0}

\newcommand{\ourabr}{{\color{black}            {LiveHand}}}

\newcommand{\MH}[1]{{\color{marcs_green}            {[MH: #1]}}}
\newcommand{\MHComment}[1]{{\color{marcs_green}            {[\textbf{MH comment: #1]}}}}
\newcommand{\AM}[1]{{\color{teal}            {[Akshay: #1]}}}
\newcommand{\ME}[1]{{\color{maroon}            {[Mohamed: #1]}}}
\newcommand{\mbr}[1]{{\color{orange}            {[MBR: #1]}}}
\newcommand{\JW}[1]{{\color{maroon}            {[Jiayi: #1]}}}
\newcommand{\todo}[1]{{\color{maroon}            {[ToDo: #1]}}}
\newcommand{\TODO}[1]{{\color{maroon}            {[ToDo: #1]}}}
\newcommand{\edits}[1]{{\color{blue}            {[New edits: #1]}}}


\newcommand{\targetimg}{\mathbf{G}}  
\newcommand{\tindex}{j} 
\newcommand{\pt}{x}  
\newcommand{\viewdir}{d}  
\newcommand{\handparam}{\psi}  
\newcommand{\handpose}{\xi}  
\newcommand{\handmesh}{\mathcal{M}}
\newcommand{\signeddist}{h}  
\newcommand{\featureop}{\mathbf{f}}  
\newcommand{\rC}{{\mathbf{C}}}
\newcommand{\mlpnw}{H_\alpha}
\newcommand{\srnw}{S_\phi}

\newcommand{\cmark}{{\color{mygreen}\ding{51}}}
\newcommand{\xmark}{{\color{myred}\ding{55}}}

\def\reta{{\textnormal{$\eta$}}}
\def\ra{{\textnormal{a}}}
\def\rb{{\textnormal{b}}}
\def\rc{{\textnormal{c}}}
\def\rd{{\textnormal{d}}}
\def\re{{\textnormal{e}}}
\def\rf{{\textnormal{f}}}
\def\rg{{\textnormal{g}}}
\def\rh{{\textnormal{h}}}
\def\ri{{\textnormal{i}}}
\def\rj{{\textnormal{j}}}
\def\rk{{\textnormal{k}}}
\def\rl{{\textnormal{l}}}
\def\rn{{\textnormal{n}}}
\def\ro{{\textnormal{o}}}
\def\rp{{\textnormal{p}}}
\def\rq{{\textnormal{q}}}
\def\rr{{\textnormal{r}}}
\def\rs{{\textnormal{s}}}
\def\rt{{\textnormal{t}}}
\def\ru{{\textnormal{u}}}
\def\rv{{\textnormal{v}}}
\def\rw{{\textnormal{w}}}
\def\rx{{\textnormal{x}}}
\def\ry{{\textnormal{y}}}
\def\rz{{\textnormal{z}}}

\def\rvepsilon{{\mathbf{\epsilon}}}
\def\rvtheta{{\mathbf{\theta}}}
\def\rva{{\mathbf{a}}}
\def\rvb{{\mathbf{b}}}
\def\rvc{{\mathbf{c}}}
\def\rvd{{\mathbf{d}}}
\def\rve{{\mathbf{e}}}
\def\rvf{{\mathbf{f}}}
\def\rvg{{\mathbf{g}}}
\def\rvh{{\mathbf{h}}}
\def\rvu{{\mathbf{i}}}
\def\rvj{{\mathbf{j}}}
\def\rvk{{\mathbf{k}}}
\def\rvl{{\mathbf{l}}}
\def\rvm{{\mathbf{m}}}
\def\rvn{{\mathbf{n}}}
\def\rvo{{\mathbf{o}}}
\def\rvp{{\mathbf{p}}}
\def\rvq{{\mathbf{q}}}
\def\rvr{{\mathbf{r}}}
\def\rvs{{\mathbf{s}}}
\def\rvt{{\mathbf{t}}}
\def\rvu{{\mathbf{u}}}
\def\rvv{{\mathbf{v}}}
\def\rvw{{\mathbf{w}}}
\def\rvx{{\mathbf{x}}}
\def\rvy{{\mathbf{y}}}
\def\rvz{{\mathbf{z}}}

\def\erva{{\textnormal{a}}}
\def\ervb{{\textnormal{b}}}
\def\ervc{{\textnormal{c}}}
\def\ervd{{\textnormal{d}}}
\def\erve{{\textnormal{e}}}
\def\ervf{{\textnormal{f}}}
\def\ervg{{\textnormal{g}}}
\def\ervh{{\textnormal{h}}}
\def\ervi{{\textnormal{i}}}
\def\ervj{{\textnormal{j}}}
\def\ervk{{\textnormal{k}}}
\def\ervl{{\textnormal{l}}}
\def\ervm{{\textnormal{m}}}
\def\ervn{{\textnormal{n}}}
\def\ervo{{\textnormal{o}}}
\def\ervp{{\textnormal{p}}}
\def\ervq{{\textnormal{q}}}
\def\ervr{{\textnormal{r}}}
\def\ervs{{\textnormal{s}}}
\def\ervt{{\textnormal{t}}}
\def\ervu{{\textnormal{u}}}
\def\ervv{{\textnormal{v}}}
\def\ervw{{\textnormal{w}}}
\def\ervx{{\textnormal{x}}}
\def\ervy{{\textnormal{y}}}
\def\ervz{{\textnormal{z}}}

\def\rmA{{\mathbf{A}}}
\def\rmB{{\mathbf{B}}}
\def\rmC{{\mathbf{C}}}
\def\rmD{{\mathbf{D}}}
\def\rmE{{\mathbf{E}}}
\def\rmF{{\mathbf{F}}}
\def\rmG{{\mathbf{G}}}
\def\rmH{{\mathbf{H}}}
\def\rmI{{\mathbf{I}}}
\def\rmJ{{\mathbf{J}}}
\def\rmK{{\mathbf{K}}}
\def\rmL{{\mathbf{L}}}
\def\rmM{{\mathbf{M}}}
\def\rmN{{\mathbf{N}}}
\def\rmO{{\mathbf{O}}}
\def\rmP{{\mathbf{P}}}
\def\rmQ{{\mathbf{Q}}}
\def\rmR{{\mathbf{R}}}
\def\rmS{{\mathbf{S}}}
\def\rmT{{\mathbf{T}}}
\def\rmU{{\mathbf{U}}}
\def\rmV{{\mathbf{V}}}
\def\rmW{{\mathbf{W}}}
\def\rmX{{\mathbf{X}}}
\def\rmY{{\mathbf{Y}}}
\def\rmZ{{\mathbf{Z}}}

\def\ermA{{\textnormal{A}}}
\def\ermB{{\textnormal{B}}}
\def\ermC{{\textnormal{C}}}
\def\ermD{{\textnormal{D}}}
\def\ermE{{\textnormal{E}}}
\def\ermF{{\textnormal{F}}}
\def\ermG{{\textnormal{G}}}
\def\ermH{{\textnormal{H}}}
\def\ermI{{\textnormal{I}}}
\def\ermJ{{\textnormal{J}}}
\def\ermK{{\textnormal{K}}}
\def\ermL{{\textnormal{L}}}
\def\ermM{{\textnormal{M}}}
\def\ermN{{\textnormal{N}}}
\def\ermO{{\textnormal{O}}}
\def\ermP{{\textnormal{P}}}
\def\ermQ{{\textnormal{Q}}}
\def\ermR{{\textnormal{R}}}
\def\ermS{{\textnormal{S}}}
\def\ermT{{\textnormal{T}}}
\def\ermU{{\textnormal{U}}}
\def\ermV{{\textnormal{V}}}
\def\ermW{{\textnormal{W}}}
\def\ermX{{\textnormal{X}}}
\def\ermY{{\textnormal{Y}}}
\def\ermZ{{\textnormal{Z}}}

\def\vzero{{\bm{0}}}
\def\vone{{\bm{1}}}
\def\vmu{{\bm{\mu}}}
\def\vtheta{{\bm{\theta}}}
\def\va{{\bm{a}}}
\def\vb{{\bm{b}}}
\def\vc{{\bm{c}}}
\def\vd{{\bm{d}}}
\def\ve{{\bm{e}}}
\def\vf{{\bm{f}}}
\def\vg{{\bm{g}}}
\def\vh{{\bm{h}}}
\def\vi{{\bm{i}}}
\def\vj{{\bm{j}}}
\def\vk{{\bm{k}}}
\def\vl{{\bm{l}}}
\def\vm{{\bm{m}}}
\def\vn{{\bm{n}}}
\def\vo{{\bm{o}}}
\def\vp{{\bm{p}}}
\def\vq{{\bm{q}}}
\def\vr{{\bm{r}}}
\def\vs{{\bm{s}}}
\def\vt{{\bm{t}}}
\def\vu{{\bm{u}}}
\def\vv{{\bm{v}}}
\def\vw{{\bm{w}}}
\def\vx{{\bm{x}}}
\def\vy{{\bm{y}}}
\def\vz{{\bm{z}}}

\title{LiveHand: Real-time and Photorealistic Neural Hand Rendering}

\author{
Akshay Mundra \textsuperscript{\rm 1,2},
\quad Mallikarjun B R \textsuperscript{\rm 1}, 
\quad Jiayi Wang \textsuperscript{\rm 1}, \\ 
\quad Marc Habermann \textsuperscript{\rm 1}, 
\quad Christian Theobalt \textsuperscript{\rm 1,2}, 
\quad Mohamed Elgharib \textsuperscript{\rm 1} \\
\small{$^{1}$ Max Planck Institute for Informatics \quad}
\small{$^{2}$ Saarland University} \\
\small{\texttt{mundra.akshay15@gmail.com, \{mbr,jwang,mhaberma,theobalt,elgharib\}@mpi-inf.mpg.de}}
}

\twocolumn[{%
\renewcommand\twocolumn[1][]{#1}%
\maketitle

\ificcvfinal\thispagestyle{empty}\fi
  \begin{center}
\centering
  \includegraphics[width=1.0\textwidth]{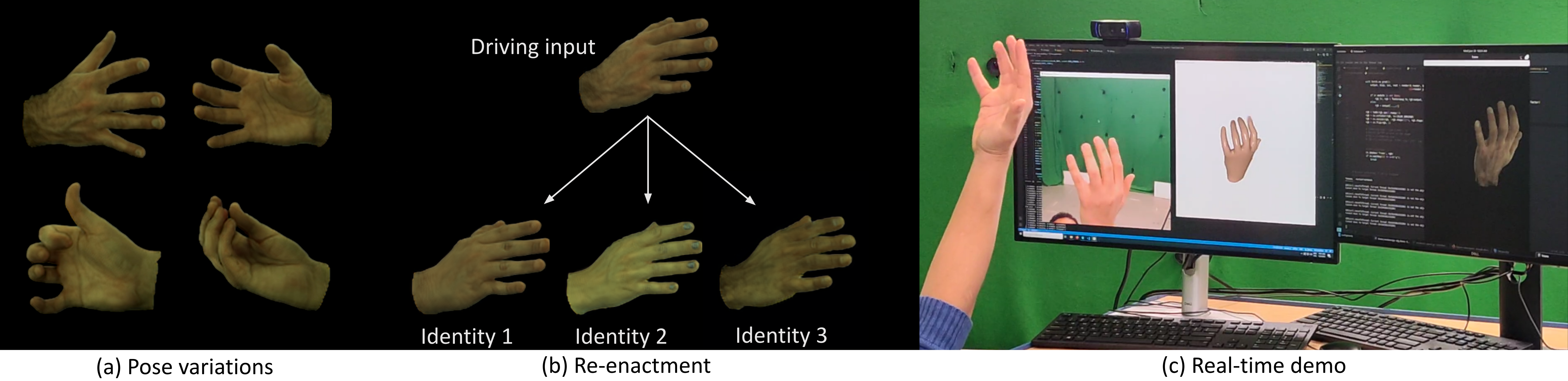}
  \captionof{figure}{We present LiveHand, the first neural implicit approach for rendering articulated hands in real-time. 
  (a) Our method captures pose-dependent effects such as hand shadows, popping veins, and skin wrinkles.
  (b) We can use the hand-pose obtained from an input sequence to re-enact different identities.
  (c) Our method is designed to optimize rendering speed and quality -- we show a live demo where we track the 3D hand-pose and render a photo-realistic hand avatar, all in real-time.}
  \label{fig:teaser}
\end{center}
}]

\begin{abstract}
    \vspace*{-0.5cm}
The human hand is the main medium through which we interact with our surroundings, making its digitization an important problem. 
While there are several works modeling the geometry of hands, little attention has been paid to capturing photo-realistic appearance. 
Moreover, for applications in extended reality and gaming, real-time rendering is critical. 
We present the first neural-implicit approach to photo-realistically render hands in real-time. 
This is a challenging problem as hands are textured and undergo strong articulations with pose-dependent effects.
However, we show that this aim is achievable through our carefully designed method. 
This includes training on a low-resolution rendering of a neural radiance field, together with a 3D-consistent super-resolution module and mesh-guided sampling and space canonicalization. 
We demonstrate a novel application of perceptual loss on the image space, which is critical for learning details accurately. 
%
%
We also show a live demo where we photo-realistically render the human hand in real-time for the first time, while also modeling pose- and view-dependent appearance effects.
We ablate all our design choices and show that they optimize for rendering speed and quality. 
Video results and our code can be accessed from \href{https://vcai.mpi-inf.mpg.de/projects/LiveHand/}{https://vcai.mpi-inf.mpg.de/projects/LiveHand/}
\end{abstract}

%
%
\section{Introduction} \label{sec:introduction}
%
%
As the popularity of VR/AR technology rises, providing a natural interface with these digital contents becomes vital. 
Undoubtedly, hands are the most intuitive mode of interaction for users in a 3D environment. 
Therefore, it is quintessential to digitize the users' hands to render their personalized, controllable, and photorealistic counterparts in the virtual world.
Achieving this is a challenging task since hand appearance is a complex function varying with both pose and viewing direction.
Moreover, ensuring real-time performance of such a system is key to enabling applications such as telepresence, teleoperation, and computer-aided design. 
%
%
\par
While the creation of photorealistic hand models is possible to some extent using traditional computer graphics techniques, it typically requires extensive manual efforts from experienced artists.
Therefore, recent research has started to investigate whether hand models can be directly derived from 2D imagery.
Here, most existing methods use some data-driven explicit model to constrain the hand geometry and appearance to a low dimensional space for the sake of tractability and robustness to occlusions \cite{MANO:SIGGRAPHASIA:2017, qian2020html, moon2020deephandmesh, li21piano, li2022nimble}.
Reconstruction is then formulated as a search in this space for the best fitting parameters.
Although these approaches can rapidly provide plausible results, the reconstruction is constrained to the space spanned by the registered hand mesh data used to create the model, thus limiting the visual quality and level of personalization.
%
%
\par
More recently, neural implicit representations~\cite{mildenhall2020nerf} have shown impressive results on static scenes for novel-view synthesis.
Some works have extended these formulations beyond static scenes to enable photorealistic renderings of articulated objects such as the human body~\cite{su2021nerf, peng2021neural, 2021narf, liu2021neural, peng2021animatable, yang2022banmo, habermann2021, habermann2022hdhumans}.
Despite their successes, very little work has been done applying these ideas to hands.
In contrast to bodies, hand motions exhibit more severe self-occlusions and more self-contact, which hinders the learning of scene representation that is consistent across different articulations.
One particular work of interest is LISA \cite{corona2022lisa}, which proposed a method to create neural hand avatars. 
Although their approach shows promising results, it does not support real-time rendering during inference and the results lack high-frequency details.
%
%
\par 
In this paper, we propose the first method for creating a \textit{photorealistic} neural hand avatar, which achieves \textit{real-time} performance while being solely learned from segmented multi-view videos of an articulated hand and respective hand pose annotations (see Fig.~\ref{fig:teaser}).
To this end, we introduce a hybrid hand model representation using the MANO hand model as a coarse proxy, which is surrounded by a neural radiance field. 
The idea is to simplify the learning problem by bounding the learnable volume through the canonicalization of global coordinates into a texture cube.
These normalized coordinates can then be fed into a shallow coordinate-based MLP to regress the scene color and density.
This formulation can also leverage the coarse mesh proxy for more efficient sampling of a low-resolution NeRF representation of the scene;
we show that this, when combined with a CNN-based super-resolution module carefully designed for efficient upsampling, can achieve real-time performance.
Moreover, we found that our highly efficient representation allows training not only on a few ray samples per iteration but on full images. 
Therefore, we can for the first time supervise an implicit scene representation using a perceptual loss on \textit{full images} during training.
Again our experiments show that this greatly improves our results over the baseline, which runs perceptual supervision on a patch basis.
Together, these design choices allow us to render and re-enact photo-realistic hands in real-time detailed enough to capture even pose- and view-dependent appearance changes.
%
%

In summary, our contributions are:
\begin{itemize}
    \item{We propose \ourabr, the first method for real-time photorealistic neural hand rendering.}
    \item{The real-time performance is achieved with our careful combination of design choices, namely, a mesh-guided 3D sampling strategy, a low-resolution neural radiance field, and a 3D-consistent super-resolution module.}
    \item{With these computationally-efficient design choices, we for the first time demonstrate that a perceptual loss on the full image can be effectively used for supervising implicit representations and that it out-performs the commonly used patch-based loss.} 
\end{itemize}
Our results demonstrate that we clearly outperform the state of the art in terms of visual quality and runtime performance.
Moreover, we show a live demo of our approach, which convincingly shows the straightforward use of our method in daily life scenarios. 
%

%
%
%
%

\begin{table}
	\begin{center}
        \resizebox{\columnwidth}{!}{
		\begin{tabular}{|c|c|c|c|c|}
			%
			\hline
			Methods                              & Real-                     & Photo-          & Pose-dep.              &  View-dep.             \\
			                                     & time                      & real            & app.                   &  app.                  \\
			\hline
   		    HTML \cite{qian2020html}  		         & \cmark                    & \xmark          & \xmark	                & \xmark	             \\
	        NIMBLE \cite{li2022nimble}  		         & \cmark                    & \cmark	       & \xmark	                & \xmark	             \\	
            LISA \cite{corona2022lisa}  		     & \xmark              	     & \xmark          & \cmark                 & \xmark	             \\
            \textbf{Ours}  			             & \cmark                    & \cmark          & \cmark                 & \cmark                 \\
			\hline
		\end{tabular}
        }
	\end{center}
    \vspace{-0.3cm}
    \caption
    	{
            Conceptual comparison of our method with other hand-modeling approaches. 
    	}
        \label{tab:concecptual_comparison}
    \vspace{-0.3cm}
\end{table}
%
%

%
\section{Related Works} \label{sec:related}
%
%
\par 
\textbf{Geometry Modeling.}
Parametric 3D morphable models map low-dimensional control variables to deforming meshes enabling easy and efficient control of the generated geometry~\cite{FLAME:SiggraphAsia2017,SMPL-X:2019,MANO:SIGGRAPHASIA:2017,alldieck2021imghum}.
Relevant to our work, MANO~\cite{MANO:SIGGRAPHASIA:2017} learns a parametric hand model using high-resolution 3D scans, parametrizing the mesh as a function of the hand shape and pose. 
Implicit geometry modeling uses a neural network to encode the geometry as an isosurface. 
Since the learned representation is resolution-independent, it can -- in theory -- be used to retrieve meshes at arbitrarily-high resolution at inference time. 
imGHUM~\cite{alldieck2021imghum} builds a parametric full-body model comprising of detailed body, face, and hand geometry.
GraspingField~\cite{GraspingField:3DV:2020} learns a signed distance function (SDF) of hand-object interaction, which fits the MANO model onto the SDF to recover the final pose estimate. 
However, none of the existing works~\cite{MANO:SIGGRAPHASIA:2017,li21piano,moon2020deephandmesh,GraspingField:3DV:2020,karunratanakul2021halo} include a component for the hand texture.
In contrast, our goal is to model the photorealistic hand appearance in real-time.
%
%
\par 
\textbf{Geometry and Appearance Modeling.}
A few approaches extend parametric mesh by complementing it with a texture map. 
HTML~\cite{qian2020html} builds a low-dimensional hand appearance model by applying principal component analysis (PCA) to texture maps of 51 subjects. 
NIMBLE~\cite{li2022nimble} uses MRI data to learn a parametric mesh model based on the bones and muscles, and uses light-stage captures to obtain the appearance maps (including albedo, normal maps, and specular maps). 
A PCA on the various components of appearance maps gives them an appearance model. 
Since both HTML and NIMBLE use a linear model to compress the appearance variations to a low-dimensional space, their expressivity is severely limited.
For example, they lack details such as veins and colored fingernails since these are person-specific attributes.
%
%
%
%
Closest to our approach is LISA~\cite{corona2022lisa}, which models the hand shape and appearance using a neural implicit field. 
%
%
The underlying MLPs are conditioned on pose and appearance parameters, allowing pose and appearance changes at inference.
However, the reconstructions lack high-frequency details, and the approach takes about one minute to render an image at $1024\times667$ pixels. 
On the other hand, we focus on creating a digital hand avatar in a person-specific setup and show photorealistic results in real-time. 
Please refer to Tab.~\ref{tab:concecptual_comparison} for a conceptual comparison of the existing hand modeling methods.
%
%
\par 
\textbf{Other Animatable Objects.} 
The literature contains works for modeling other animatable objects such as the human face ~\cite{Lombardi21,Cao22,Zheng:CVPR:2022,grassal2022neural,Gao2022nerfblendshape,Ma21}, human body ~\cite{habermann2021,vid_based_charcs,bhatnagar2020ipnet,liu2021neural,yang2022banmo,peng2021neural,su2021nerf}, and animals ~\cite{Luo22}.
%
The face related methods can not handle large deformations~\cite{Lombardi21,Cao22,Gao2022nerfblendshape,Ma21} and/or are not real-time ~\cite{grassal2022neural,Zheng:CVPR:2022,Gao2022nerfblendshape}, while ~\cite{Luo22} does not model pose-dependent appearance effects.
To handle more articulated motions that occur in the human body, two classes of body-specific methods have been proposed. 
The explicit mesh-based methods ~\cite{vid_based_charcs,habermann2021,Bagautdinov21} rely on a template mesh obtained from a static scene and then learn appearance in the mesh space either by retrieval~\cite{vid_based_charcs} or by using a CNN to directly regress the texture map~\cite{habermann2021,Bagautdinov21}. 
However, due to the strong reliance on a template mesh, the learned appearance becomes blurry if the deformed template mesh does not match the real deformation of the surface.
%
%
In contrast, neural implicit models have the capacity to learn more fine-grained deformations at much higher resolution. For example, it has been used to model the geometry and appearance of clothed humans \cite{su2021nerf, peng2021neural, 2021narf, liu2021neural, peng2021animatable, yang2022banmo, habermann2022hdhumans, hu2021hvtr, peng2022animatable, jiang2022selfrecon}.
However, these can not operate in real-time.
Some efficient approaches ~\cite{prokudin2021smplpix, Raj_2021_CVPR} formulate the rendering task as an image-translation problem, but suffer from inaccuracies in parametric model fitting.
Yet another line of implicit body-modeling approaches \cite{Remelli22,Suo21RT,shao2022floren} require RGB images from multiple cameras at test time, and thus can not be controlled with arbitrary poses.
Extending the body modeling methods to human hands is not trivial, as hands exhibit even stronger articulation, which in turn results in severe self-occlusion and other pose-dependent effects. 
Our proposed method tackles this setting by utilizing elements from both mesh-based and neural implicit modeling to create a detailed model that runs in real-time.
%

%
%
\section{Methodology} \label{sec:methodology}
\begin{figure*}
    \centering
    \includegraphics[width=\linewidth]{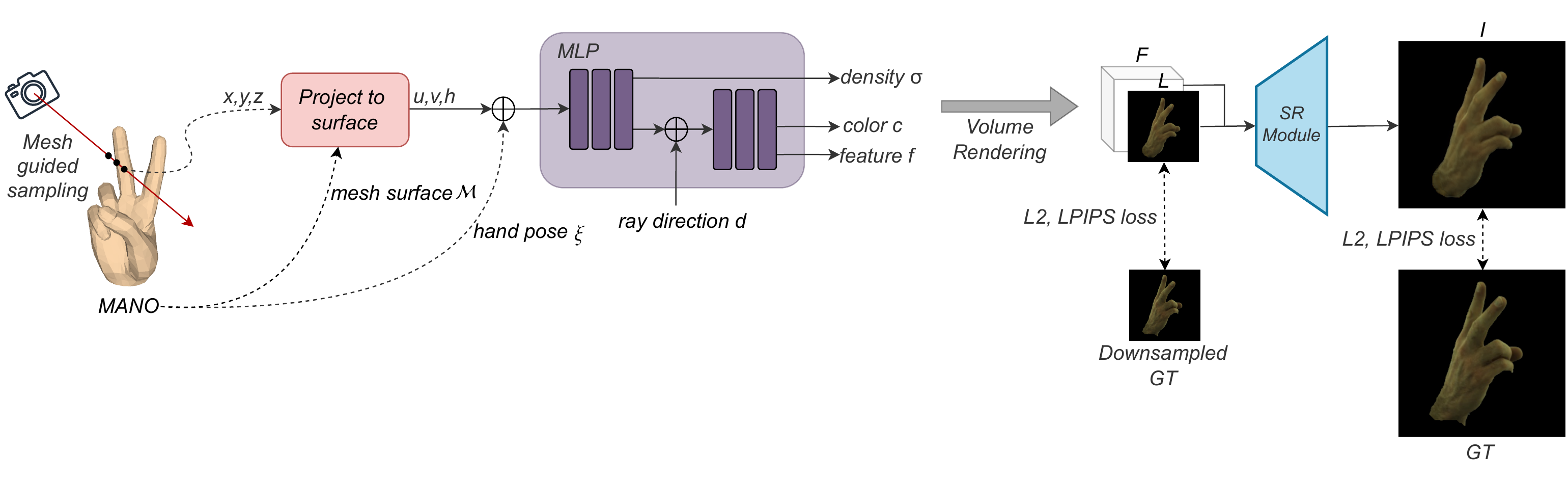} 
    \vspace{-7mm}
    \caption{
    \textbf{Overview of our approach.} 
    Given a hand pose and camera view, our method renders a photorealistic image of the hand in real-time.
    To this end, we employ an efficient MANO mesh-guided sampling and canonicalization strategy.
    The hand appearance is captured by an MLP that maps points from the canonicalized texture space to radiance values.
    We then leverage volume rendering to obtain a low-resolution image-aligned feature tensor where the first three channels contain the RGB image of the hand. 
    Finally, a super-resolution module up-samples the tensor to obtain the final full-resolution image. 
    Since our method achieves a fast inference speed, we can supervise it with a perceptual loss on the full image resolution.
    }
    \label{fig:overview}
\end{figure*}

Given multi-view images {\small $\{\targetimg_\tindex^p | j=1 \ldots N, p=1 \ldots P\}$} for $P$ 
frames captured from $N$ viewpoints and the corresponding coarse parametric hand meshes $\{\handmesh(\handparam^p)| p=1 \ldots P\}$, 
our method creates a photo-realistic hand avatar that can accurately model hand-pose and view-dependent appearance effects, and can be rendered in real-time.
An overview of our method is shown in Fig.~\ref{fig:overview}.
Given the hand parameters $\handparam$, we can canonicalize every point in the scene based on the point's projection onto the posed mesh $\handmesh(\handparam)$.
The 3D coordinates are then re-parameterized in terms of the corresponding texture coordinates after projection. 
A multi-layer perception (MLP) $\mlpnw$ is then trained to map the re-parameterized coordinates to a radiance field, conditioned on articulation parameters. 
For the given camera extrinsics and intrinsics, we render low-resolution images and image-aligned feature maps using volumetric rendering, which is then up-sampled using a super-resolution network $\srnw$ to obtain the final rendering.
In this section, we initially describe the hand model required to build the neural hand representation in Sec.~\ref{subsec:pre_process}, the scene representation in Sec.~\ref{subsec:representaion}, and its efficient $2D$ rendering in Sec.~\ref{subsec:rendering}. Finally, in Sec.~\ref{subsec:training}, we describe how our neural hand model can be effectively trained. 
%
%
%
\subsection{MANO Model} \label{subsec:pre_process}
We leverage the MANO~\cite{MANO:SIGGRAPHASIA:2017} model to parameterize the approximate hand geometry.
MANO maps the model parameter $\handparam$ to a posed mesh $\handmesh$ using its Linear Blend Skinning (LBS) weights $W$ and a canonical hand mesh $\overline{\handmesh}$.
%
\begin{equation}
    \handmesh(\handparam) = MANO(\overline{\handmesh}, \handparam, W)
\end{equation}
%
$\handparam: \{\theta, \beta, t, R\} \in \mathbb{R}^{61}$ consists of the articulation parameters $\theta \in \mathbb{R}^{45}$, shape parameters $\beta \in \mathbb{R}^{10}$, and the global translation $t \in \mathbb{R}^{3}$ and rotation in axis-angle format $R \in \mathbb{R}^{3}$.
We refer the readers to~\cite{MANO:SIGGRAPHASIA:2017} for more details.
For convenience, we also define hand pose as $\xi:\{\theta, R\} \in \mathbb{R}^{48}$ here. $\xi$ encodes only the articulation and orientation of the hand, and is, thus, independent of identity and position in global 3D space.
%
%
\subsection{Implicit Hand Representation} \label{subsec:representaion}
Inspired by the state-of-the-art implicit novel view synthesis method, NeRF~\cite{mildenhall2020nerf}, we model our hand avatar with a view-dependent implicit representation. 
Since NeRF can only capture static scenes, we must extend the radiance field to account for deformations. 
In this section, we systematically motivate and describe our chosen representation. 
%
%
\par \textbf{Naive Conditioning.}
One way to formulate the hand radiance field $\mlpnw$ is by naively conditioning it as follows:
%
\begin{equation}
\label{eqn:hand_model_naive}
   \mlpnw : (\pt, \viewdir, \handpose) \rightarrow (\mathbf{c}, \sigma)
\end{equation}
%
where $\pt$ is 3D point, $\viewdir$ is the viewing direction, $\handpose$ is the hand pose, $\mathbf{c}$ is the color and $\sigma$ is the density. 
The trainable radiance field $\mlpnw$ is parameterized by an MLP with parameters $\alpha$. 
However, this leads to poor generalization to novel test hand poses as will be shown in Sec.~\ref{sec:experiments}. 
This is because any point on the hand surface gets mapped to completely different world coordinates based on the hand pose. 
%
%
\par \textbf{Per-bone Canonicalization.}
One way to overcome this problem in the literature~\cite{corona2022lisa} is to canonicalize the scene with respect to the hand pose.
Specifically, a point in world space is transformed into each bone's local coordinate systems obtained from a skeleton pose estimate. 
Separate implicit fields are learnt in the local coordinate systems, which are combined as follows: 
%
\vspace{-2.5mm}
\begin{equation}
    \sigma = \sum_{k=1}^{n_b} w_k \sigma_k\;,\hspace{2mm}\mathbf{c} = \sum_{k=1}^{n_b} w_k \mathbf{c}_k\;
    \label{eqn:per_bone_canonicalization}
\end{equation}
%
where $w$ is analogous to LBS weights. We evaluate such a canonicalization approach in Sec.~\ref{sec:experiments}. 
Such a per-bone canonicalization requires inferring multiple MLPs for each 3D point, making it slower for both training and inference. 
%
%
\par \textbf{Mesh-based Canoncialization.}
For a more efficient representation, we take inspiration from mesh-based texturing which associates each point on the mesh surface with a 2D texture coordinate $(u,v) \in [0,1] \times [0,1]$ from which a color value can be obtained using a texture image. 
We extend this surface representation to 3D volumes by introducing a signed distance $\signeddist$ to support volume rendering and to account for the coarseness of the MANO-based geometry approximation. 
More concretely, for a given point $\pt$ in 3D, we first find its projection on the given MANO surface. The $(u,v)$ co-ordinate of this projected point can be estimated by performing barycentric interpolation on the $(u,v)$ coordinates of the corresponding mesh-triangle vertices.
The signed distance $\signeddist$ of the sampling point to its projection on the mesh is used to disambiguate points orthogonal to the mesh surface ~\cite{liu2021neural}.
With this canonicalization, we can formulate the radiance field mapping as,
%
\vspace{-2.5mm}
\begin{equation}
    \label{eqn:hand_model_canonical}
    \mlpnw : (u, v, \signeddist, \viewdir) \rightarrow (\mathbf{c}, \sigma)
\end{equation}
%
This allows us to canonicalize the world coordinates to a representation that stays consistent with respect to hand surface irrespective of hand pose $\handpose$, thus preventing the dispersion of learned features in the input space.
In practice, we apply positional encoding~\cite{mildenhall2020nerf} to all inputs in Eq.~\ref{eqn:hand_model_canonical}.

This canonicalized $uvh$ space does not contain any pose information. Since a point on the hand surface could have a different appearance based on the hand pose, we also explicitly condition our model with the hand pose $\handpose$ after canonicalization.
This leads to the modified representation:
%
\begin{equation}
\vspace{-2.5mm}
\label{eqn:hand_model_full}
    \mlpnw : (u, v, \signeddist, \viewdir, \handpose) \rightarrow (\mathbf{c}, \sigma)
\end{equation}
%
Note that although we rely on the coarse hand mesh for canonicalization, the implicit representation $\mlpnw$ can learn fine-scale details that are hard to model using MANO mesh alone.
We show this later in Sec.~\ref{sec:experiments} where our method significantly outperforms a baseline that naively textures the coarse MANO mesh using ground truth images.
%
%
\subsection{Efficient Rendering} \label{subsec:rendering}
%
%
%
Since $\mlpnw$ is parameterized with an MLP, it can be queried to regress the density $\sigma$ and color $\mathbf{c}$ for each point in 3D space. 
For a ray with origin $\rvo$ and direction $\rvd$, volumetric integration - as proposed in NeRF~\cite{mildenhall2020nerf} - can be used to obtain the integrated color $\rmC$ for the ray $\rvr(t) = \rvo + t\rvd$:
%
\begin{align}
\rmC(\rvr) &= \int_{t_n}^{t_f} T(t) \sigma(\rvr(t)) \rvc(\rvr(t)) dt \nonumber \\
\text{where} \quad & T(t) = \text{exp}(-\int_{t_n}^{t}\sigma(\rvr(s)) ds),
\label{eq:nerf_vol_int}
\end{align}
and $t_n$ and $t_f$ are near and far bounds.
This integral can be approximated through stratified sampling within the bounds. 
However, such a strategy will waste samples on regions that do not contain useful features. 
Hierarchical sampling was introduced in NeRF~\cite{mildenhall2020nerf} to address this inefficiency.
However, this involves the use of two MLPs to encode both the coarse and detailed scene, and sampling the scene twice.

\textbf{Mesh-Guided Sampling.}
To make the rendering faster, we utilize the coarse MANO geometry to efficiently sample points around the approximate hand surface.
Specifically, to define the bounds of each ray, we use the depth rendering of the coarse mesh to constrain the samples to lie close to the approximate surface \cite{habermann2022hdhumans}.
This eliminates the two-pass approach needed for hierarchical sampling.
%
%

\textbf{Super-resolution.}
Although this efficient sampling strategy improves the run-time, it still can not achieve real-time rendering speeds.
We introduce a super-resolution network~\cite{Chan2022EG3d} $\srnw$ that can super-resolve the rendered output in a 3D consistent manner. 
To do so, we first modify the $\mlpnw$ to additionally predict a $29$-channel $\featureop$, which encodes scene features alongside the color to capture additional details. 
We accomplish this by extending Eq.~\ref{eqn:hand_model_canonical} with:
%
\vspace{-1mm}
\begin{equation}
\label{eqn:hand_model_feature_output}
    \mlpnw : (u, v, \signeddist, \viewdir, \handpose) \rightarrow (\mathbf{c}, \featureop, \sigma)
\end{equation}
%
We then apply volumetric integration as done in Eq.~\ref{eq:nerf_vol_int} to obtain low-resolution renderings of color $L_j^p$ and features $F_j^p$ for each viewpoint $j$ and hand pose $p$.
\par 
These low-resolution encodings are used in a super-resolution module
%
\vspace{-2.5mm}
\begin{equation}
    \srnw : (L, F) \rightarrow I
\end{equation}
%
to recover a high-resolution image $I_j^p$ that preserves the details. 
To ensure efficiency, we parameterize $\srnw$ using a CNN-based network with the trainable parameters $\phi$.
%
%
\subsection{Training} \label{subsec:training}
As described in the previous section, we need to learn the parameters of the MLP $\mlpnw$ and super-resolution module $\srnw$ using the multi-view image sequence. 
%
%
\par \textbf{Color Calibration.}
As multi-view images, in general, are not color corrected to be consistent across views, we compensate for this, as done in Neural Volumes~\cite{Lombardi:2019}, by learning separate per-camera gain and bias parameters $g_j$ and $b_j$.
%
%
\par \textbf{Objective Function.}
We train the parameters of our modules $\mlpnw$ and $\srnw$ in a supervised manner using the following loss functions 
%
\begin{equation}
    \mathcal{L} = \mathcal{L}_{rec} + \mathcal{L}_{perc}
\end{equation}
%
between ground truth target image $\targetimg_\tindex^p$ and rendering image $I_j^p$ using gradient descent.
Here $\mathcal{L}_{rec}$ is the L2 reconstruction loss given by:
%
\begin{equation}
    \mathcal{L}_{rec} = || G^p_j - I^p_j(\alpha, \phi)||_2
\end{equation}
%
To capture the perceptual difference in the image, we apply $\mathcal{L}_{perc}$ as suggested in~\cite{Zhang_2018_CVPR} \
%
\begin{equation}
    \mathcal{L}_{perc} = || f(G^p_j) - f(I^p_j(\alpha, \phi))||_2
\end{equation}
%
Where $f(\cdot)$ is the activation of the $conv1$-$conv5$ layers in pre-trained VGG network~\cite{simonyan2014very}.  
Thanks to our efficient design choices, we can apply the perceptual loss on the full image, as opposed to the traditional approach of applying it on smaller patches \cite{weng_humannerf_2022_cvpr}. 
We show later that the perceptual loss plays a vital role in recovering high-frequency details, and our image-based approach improves photorealism over using the patch-based strategy (see Tab.~\ref{tab:ablation}, Fig.~\ref{fig:ablation} and the supplementary video).
We employ the above loss functions to both low-resolution volumetrically rendered images and high-resolution super-resolved images.
%
%
%

\section{Experiments} \label{sec:experiments}

We use the publicly released version of the InterHand2.6M benchmark for our experiments.
The dataset contains multi-view sequences of different users performing a wide range of actions at 5 FPS and $512\times334$ pixels resolution. 
To test our method, we select the right-hand sequences from four users in the ``train/capture0", ``train/capture5", ``test/capture0", and ``test/capture1" subsets. 
We reserve the last 50 frames of each capture for evaluation and use the rest for training. 

We show that the advantages of our proposed model work synergistically together to enable the first demo for real-time photorealistic neural hand reenactment.
The details of this demo and its results are presented in Section~\ref{subsec:demo}.
We additionally provide quantitative and qualitative evaluations of our method on the established benchmark in Section~\ref{subsec:sota} and Section~\ref{subsec:ablation}.
For this, we used PSNR, LPIPS, and FID metrics for numerical evaluation.
Following the conventions of~\cite{Zhang_2018_CVPR}, LPIPS score is calculated using AlexNet backbone.
For rendering speed, we report the time it takes to render an image on an NVIDIA GeForce RTX 3090 at the training resolution (i.e. $512\times334$ pixels) in frames per second (FPS).
For super-resolution experiments, volumetric integration produces a rendering at $256\times167$ pixels which are then super-resolved to $512\times334$ pixels. 
More implementation details and results can be found in the supplementary material.
%

\subsection{Application: Real-time Hand Reenactment}
\label{subsec:demo}

\begin{figure*}[]
	\centering
    \includegraphics[width=2\columnwidth]{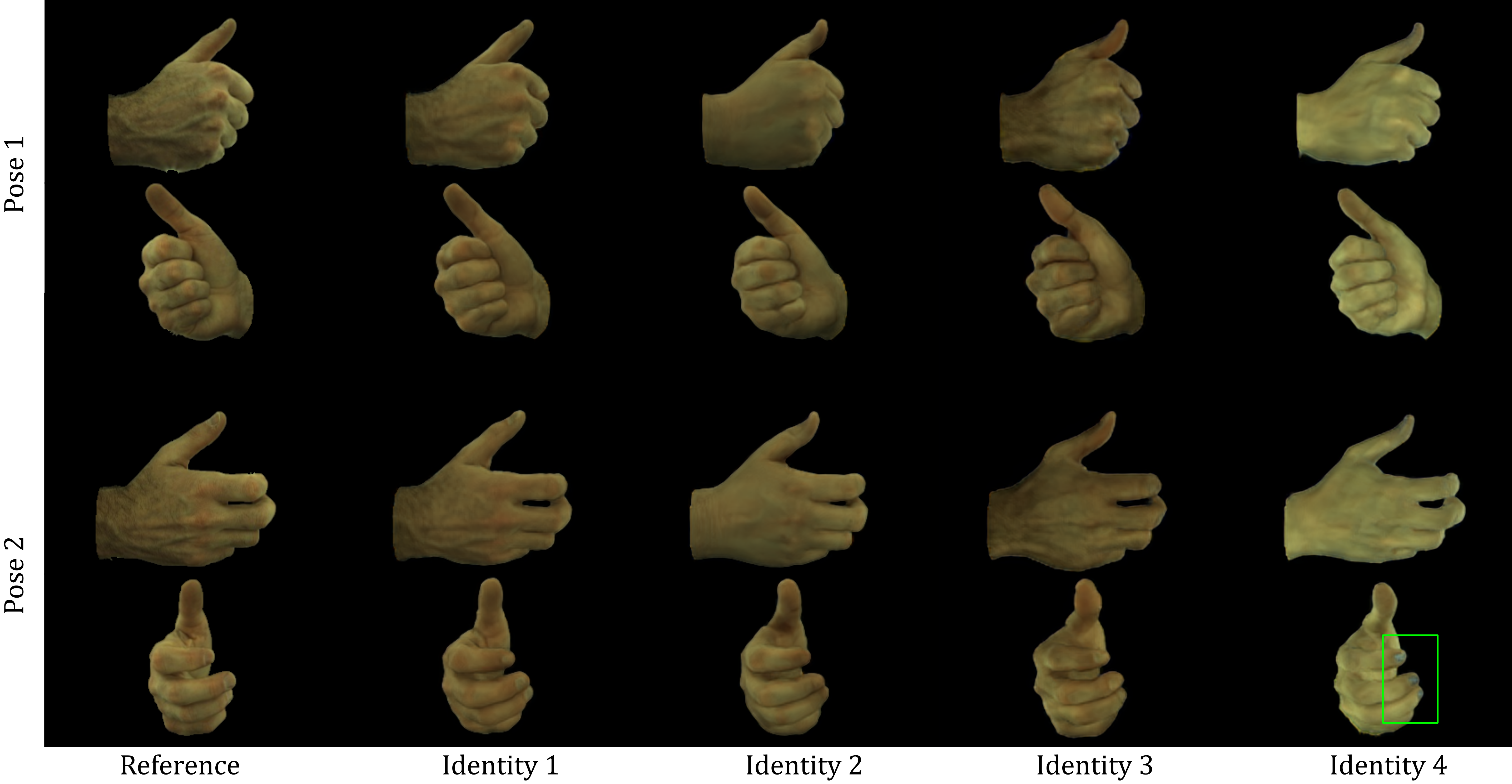}
	\caption{
 \textbf{Hand Reenactment.} 
 Our method can transfer the pose of a reference actor (Reference) to new identities (Identity 1-4). 
 Note that our model captures pose-dependent changes, which is especially apparent for veins and in the knuckle region.
 It also captures view-dependent shading and self-shadowing effects.}
	\label{fig:qualitative}
\end{figure*}

\begin{figure}[]
    \centering
    


    \begin{subfigure}{0.3\columnwidth}
        \includegraphics[width=\columnwidth,trim={9cm 8cm 38cm 10cm},clip]{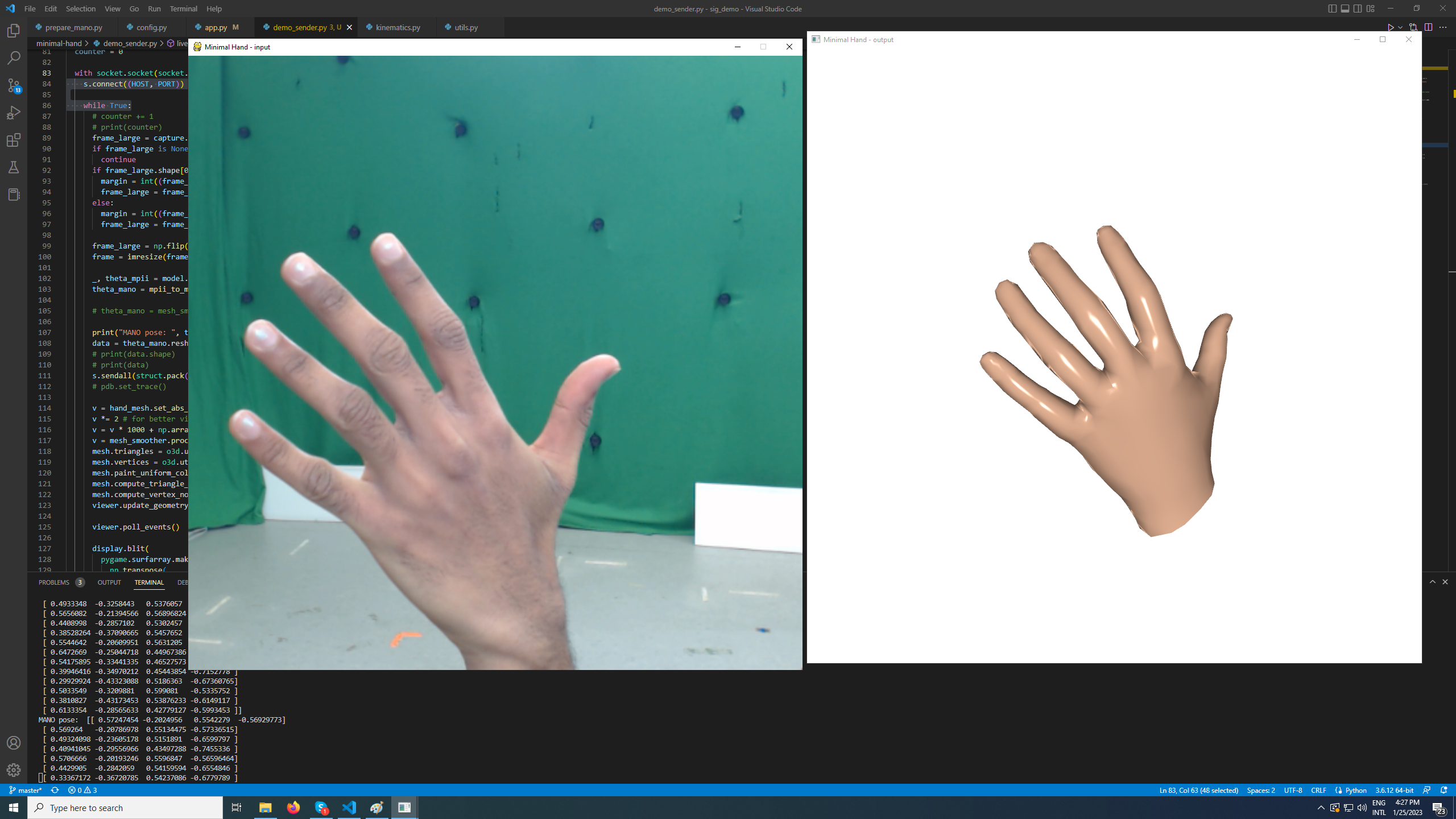}
        \caption{Driving input}
    \end{subfigure}
    \begin{subfigure}{0.3\columnwidth}
        \includegraphics[width=\columnwidth,trim={45cm 14cm 8cm 10cm},clip]{figures/demo_tracker2.png}
        \caption{Estimated pose}
    \end{subfigure}
    \begin{subfigure}{0.3\columnwidth}
        \includegraphics[width=\columnwidth,trim={26cm 12cm 24cm 8cm},clip]{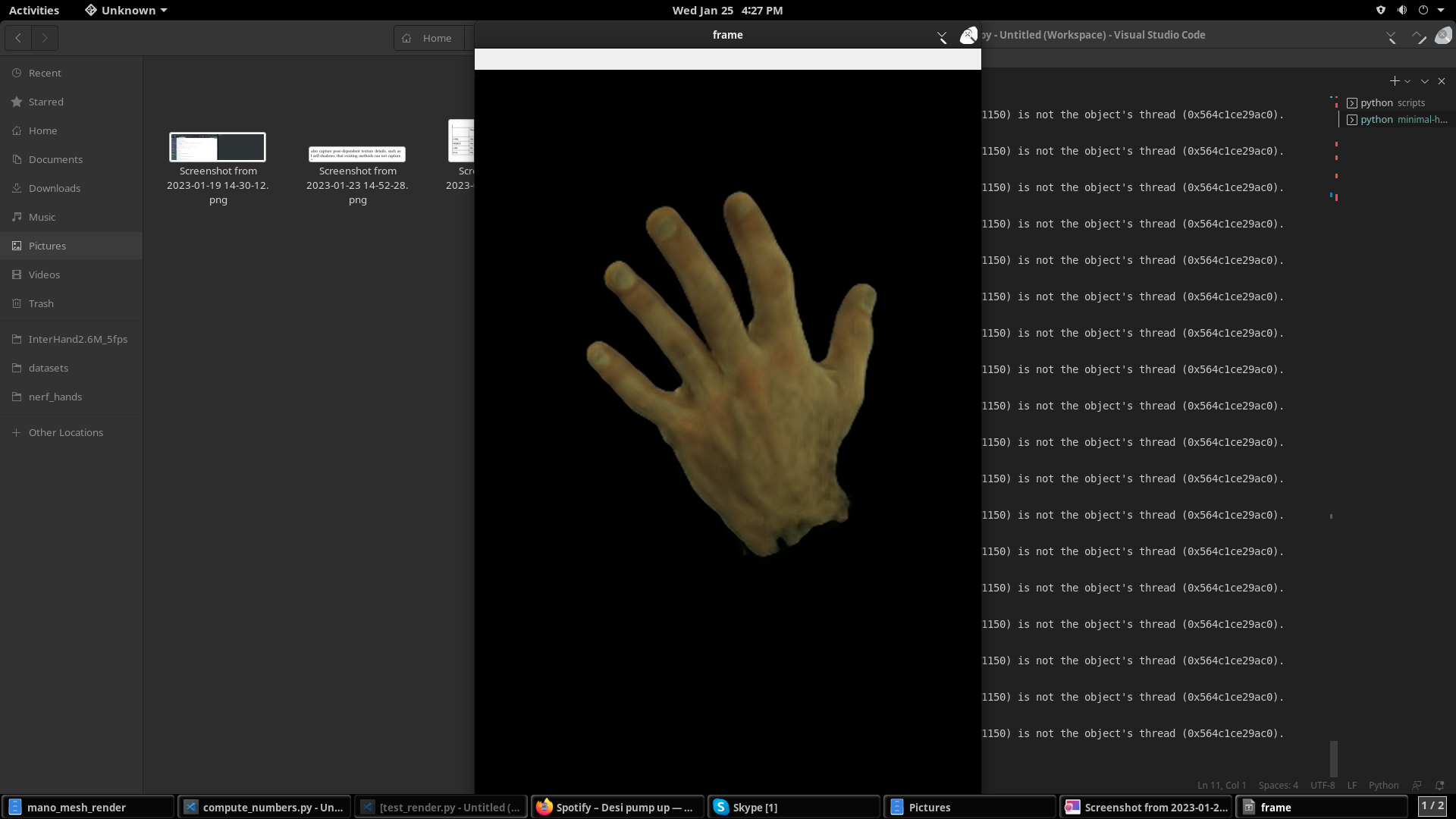}
        \caption{Our rendering}
    \end{subfigure}
    
    \caption{\textbf{Demo Visualization.} The real-time demo takes in a monocular RGB input (left) to estimate the MANO parameters (center). The MANO pose is then transferred to the target identity using our method (right).}
    \label{fig:demo}
\end{figure}

We carefully design our method specifically for real-time hand reenactment applications. 
After training our neural implicit representation $\mlpnw$ and the super-resolution module $\srnw$ to create a user's hand avatar, we can drive the articulation of that hand using new motion.
Fig.~\ref{fig:qualitative} show this transfer of hand performance from a reference user (`Reference') to $4$ learned identities.
Note that our approach is able to generalize well across identities even when the driving poses were not seen during training.
Note how the avatar of each identity captures high-frequency skin texture as well as hand-pose dependent illumination, which contributes to the photo-realism of our renderings.

To show that this method can work in real applications, we also implement a live demo. 
This application consists of two parts: a hand tracker which estimates a posed MANO mesh, and a hand avatar trained using our methods on InterHand2.6M. 
We estimate the pose using ~\cite{zhou2020monocular} and pass it to our method for rendering.
%
%
%
The pose estimator takes $10$ milliseconds while rendering our hand avatar takes $20$ milliseconds on average, giving our system an effective speed of $33$ FPS. 
We show the qualitative results of this demo in Fig.~\ref{fig:demo}.
Note the plausible high-frequency details of the rendered hand avatar driven by new poses captured live in the monocular RGB stream.
We encourage the readers to check the supplementary video for the demo, as well as 3D consistent rendering sequences with view-dependent effects.

\subsection{Comparison to State of the Art}\label{subsec:sota}

The only other neural implicit hand model that exists in the literature is LISA ~\cite{corona2022lisa}. 
As their method is trained and evaluated on an unreleased high-resolution version of the Interhand2.6M dataset and the code is not publicly available, we re-implemented their approach for a fair comparison. 
As an additional baseline, we use the body modeling method A-NeRF ~\cite{su2021nerf} and adapt it for hand modeling. 
We also compare against SMPLpix ~\cite{prokudin2021smplpix} because of its real-time performance. We adapt it to hands by changing the conditioning input from SMPL to MANO renderings.
Because our method requires a coarse hand mesh for canonicalization, we also compare against a baseline explicit method that re-textures this mesh using a pre-estimated texture map (`Mesh wrapping').
For this, we extract the texture from a flat-hand pose and wrap it to the target poses.
\begin{table}[] 
\centering
\resizebox{\columnwidth}{!}{
\begin{tabular}{@{}lccccc@{}}
\toprule
   & PSNR $\uparrow$  & LPIPS(x1000) $\downarrow$ & FID $\downarrow$   & FPS $\uparrow$ 
\\
 \cmidrule(lr){2-5}
 \text{Mesh wrapping}                  &    28.28    &   49.44    &  298.28   &    \textbf{82.33}
 \\ 
 \text{SMPLpix ~\cite{prokudin2021smplpix}}                  &    \textbf{32.37}    &   26.57    &  202.99   &    58.82
 \\
 \text{A-NeRF* ~\cite{su2021nerf}}         &    28.07    &   94.41   &   318.61    &    0.83
 \\
 \text{LISA* ~\cite{corona2022lisa}}      &     29.36    &   78.46  &   255.43   &      3.70
 \\
 \text{\textbf{Ours}}       &      32.04     &   \textbf{25.73}   &   \textbf{197.39}    &    45.45
 \\
\bottomrule
\end{tabular}
}
\caption{
\textbf{Comparison on InterHand2.6M ~\cite{Moon_2020_ECCV_InterHand2.6M}.} * indicates we use our implementation of the approach. 
}
\label{tab:sota} 
\end{table} 
\begin{figure*}[]
	\centering
    \includegraphics[width=2\columnwidth]{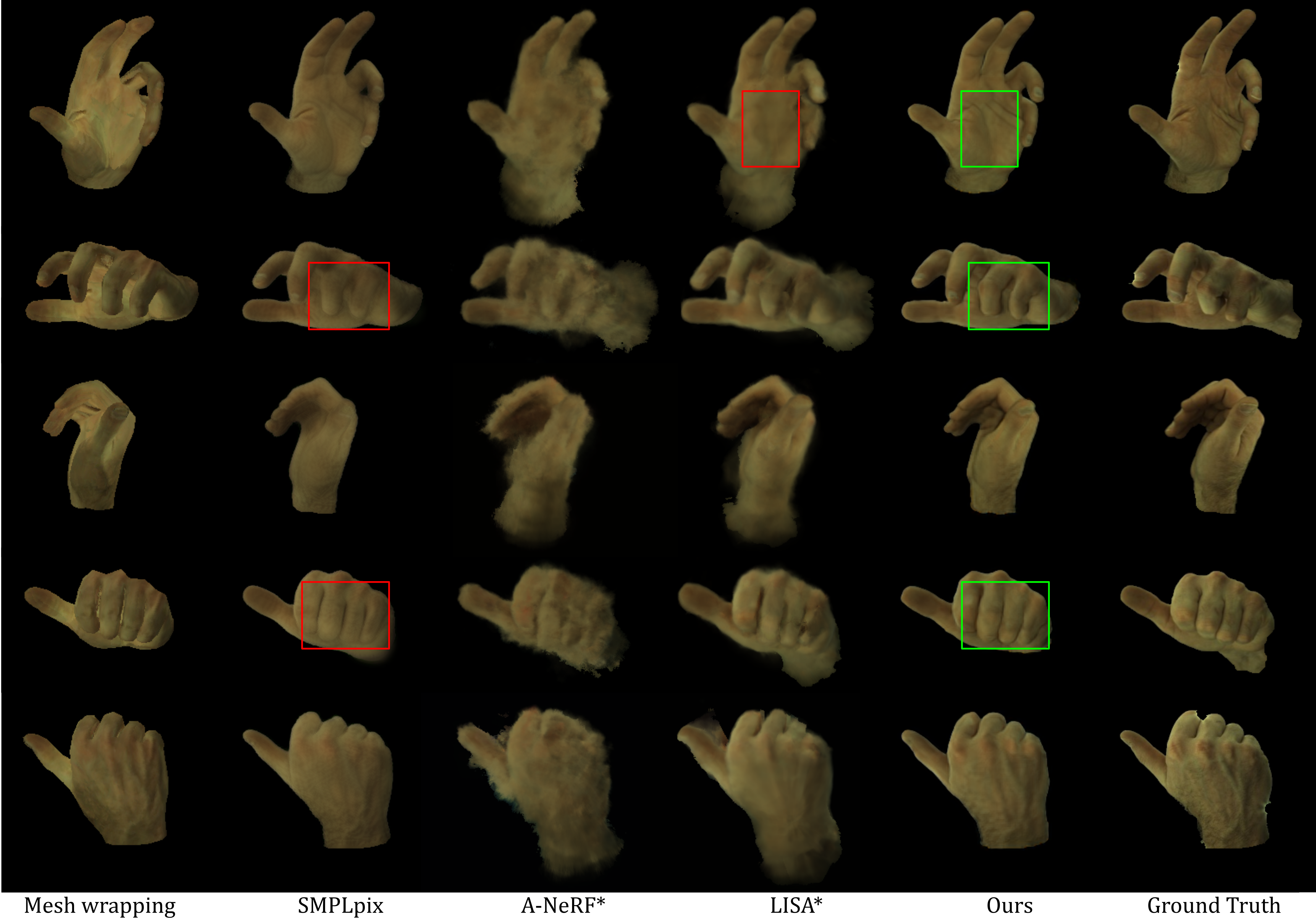}
	\caption{\textbf{Comparison to SoTA on unseen hand poses.} A-NeRF and Mesh wrapping produce artifacts while SMPLpix and LISA do not capture high-frequency details. Our method outperforms these approaches and captures high-frequency details.}
	\label{fig:sota}
    \vspace{-0.2cm}
\end{figure*}

As shown in Table~\ref{tab:sota},  our method outperforms other neural implicit baselines while also being real-time.
These improvements in the metrics also translate to significant improvements in perceptual quality on the test set, which can be seen in Fig.~\ref{fig:sota}.
We hypothesize that this is owing to our improved canonicalization strategy and our use of perceptual loss. 
Both A-NeRF and LISA use per-part canonicalization similar to the one described in Eq.\ref{eqn:per_bone_canonicalization}. 
However, learning to combine per-part output is not trivial, and could lead the ambiguities in case of severe articulations.
Moreover, as we will show in Sec.~\ref{subsec:ablation}, our addition of a perceptual loss drastically improves the level of detail the model can capture over those obtained from simple per-pixel loss used in A-NeRF and LISA.

SMPLpix comes close to our method quantitatively, but fails to capture the details, as shown in Fig.~\ref{fig:sota}. This is because, unlike our method, SMPLpix can not account for person-specific geometric changes as it strictly relies on coarse MANO geometry.

Our method also significantly outperforms the mesh wrapping baseline, quantitatively and qualitatively.
Note that modern graphics pipelines can achieve much higher frame rates for mesh rendering based on their implementation, and we only benchmark ours. 
But by no means can such a simple rendering achieve the complex appearance effects and photorealism as our method can.
This demonstrates that our model can learn improvements upon what is possible using only the coarse geometric initialization.

We show additional comparisons using a synthetic dataset in the supplementary document.

\subsection{Ablation}\label{subsec:ablation}

\begin{table}[] 
\centering
\resizebox{\columnwidth}{!}{
\begin{tabular}{@{}lccccc@{}}
\toprule
   & PSNR $\uparrow$ & LPIPS(x1000) $\downarrow$ &   FID  $\downarrow$ &  \#parameters $\downarrow$ & FPS $\uparrow$ 
\\
 \cmidrule(lr){2-6}
 xyz  &   29.31  &  42.50   &   247.77    &    0.95M      &    43.03
 \\
 per-bone xyz &  \textbf{32.51}  &   \textbf{23.82}    &   198.95   &  1.14M     &    27.04
 \\
 uvh w.o. pose cond. &   30.33   &   32.36   &  204.24  &   \textbf{0.40M}       &   \textbf{45.73}
 \\
 \textbf{Ours} (uvh w. pose cond.) &   32.04   &   25.73   &  \textbf{197.39}  &    0.41M      &    45.45
 \\
\bottomrule
\end{tabular}
}
\caption{
\textbf{Ablation on various canonicalization strategies.} Our approach optimizes for both quality and speed.
} 
\label{tab:canonicalization} 
\vspace{-0.2cm}
\end{table} 

\begin{figure*}[]
	\centering
    \includegraphics[width=2\columnwidth]{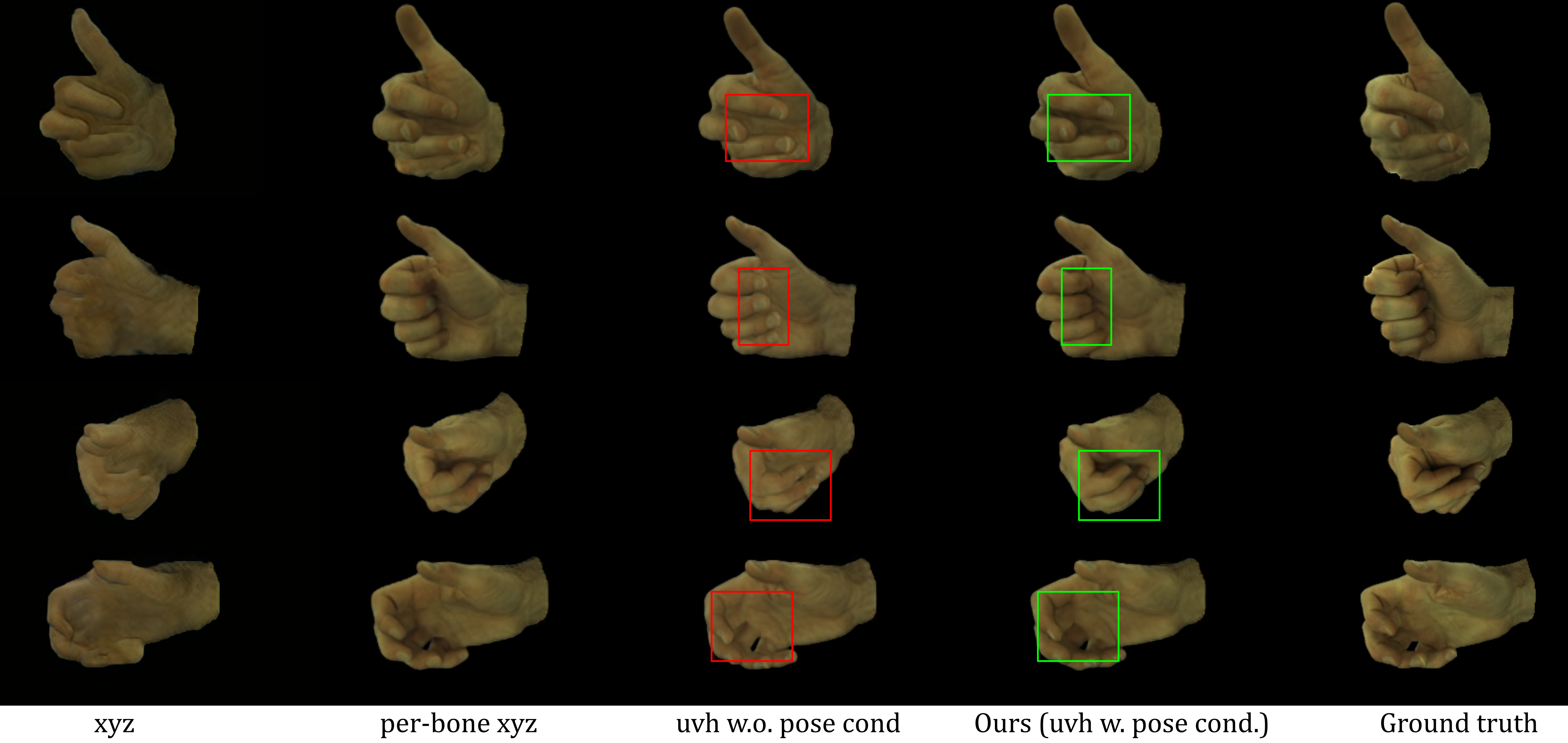}
	\caption{\textbf{Canonicalization Ablation.} 
 Global xyz coordinates with naive conditioning fails to generalize to novel poses. 
 Our proposed uvh canonicalization achieves similar visual results to per-bone xyz canonicalization while being much faster. 
 Note that hand pose conditioning is vital for capturing pose-dependent effects such as self-shadowing (see red and green regions).}
\label{fig:ablation_canonicalisation}
\end{figure*}

Our design choices are crucial for optimizing both the rendering quality and processing speed. 
To evaluate their significance, we perform ablation studies of various components. 
First, we report the impact of different canonicalization strategies on the metrics in Tab.~\ref{tab:canonicalization} and on visual quality in Fig.~\ref{fig:ablation_canonicalisation}. 
We see that naive pose conditioning (`xyz') performs the worse in all metrics, and the results are blurry and indistinct. 
While per-bone canonicalization (`per-bone xyz') produces high-quality renderings, our formulation is $1.7$ times faster as it does not rely on the evaluation of multiple MLPs.
Finally, our experiments show that without pose conditioning (`uvh w.o. pose cond.'), the performance of our method drops as it is vital for capturing pose-dependent effects such as self-shadowing and skin wrinkles, and this can be seen in Fig.~\ref{fig:ablation_canonicalisation}.

\begin{table}[]
\centering
\resizebox{\columnwidth}{!}{

\begin{tabular}{@{}llcccc@{}}
\toprule
   &   &   PSNR $\uparrow$ & LPIPS(x1000) $\downarrow$ & FID $\downarrow$ & FPS $\uparrow$ 
\\
 \cmidrule(lr){3-6}
\multicolumn{2}{l}{w.o. mesh-guided samp.}           &     31.25    &   25.95    &    202.40    &    9.07   \\
\multicolumn{1}{c|}{}   & w.o. $\mathcal{L}_{perc}$              &   \textbf{32.69}    &   38.45   &  226.78   &    19.64   \\
\multicolumn{1}{c|}{w.o. SR} & patch $\mathcal{L}_{perc}$      &   30.52   &    31.13   &   197.70   &   19.52  \\
\multicolumn{1}{c|}{}   & full $\mathcal{L}_{perc}$           &  31.61   &   26.63    &  \textbf{197.35}  &   19.37   \\ 
\multicolumn{2}{l}{\textbf{Ours} (full $\mathcal{L}_{perc}$)}                     &       32.04    &   \textbf{25.73}   &   197.39    &    \textbf{45.45}
 \\
\bottomrule
\end{tabular}
}
\caption{
\textbf{Ablation study on model components.} All design choices consistently improve the accuracy and runtime.
} 
\label{tab:ablation} 
\vspace{-0.2cm}
\end{table} 
\begin{figure*}[]
	\centering
    \includegraphics[width=2\columnwidth]{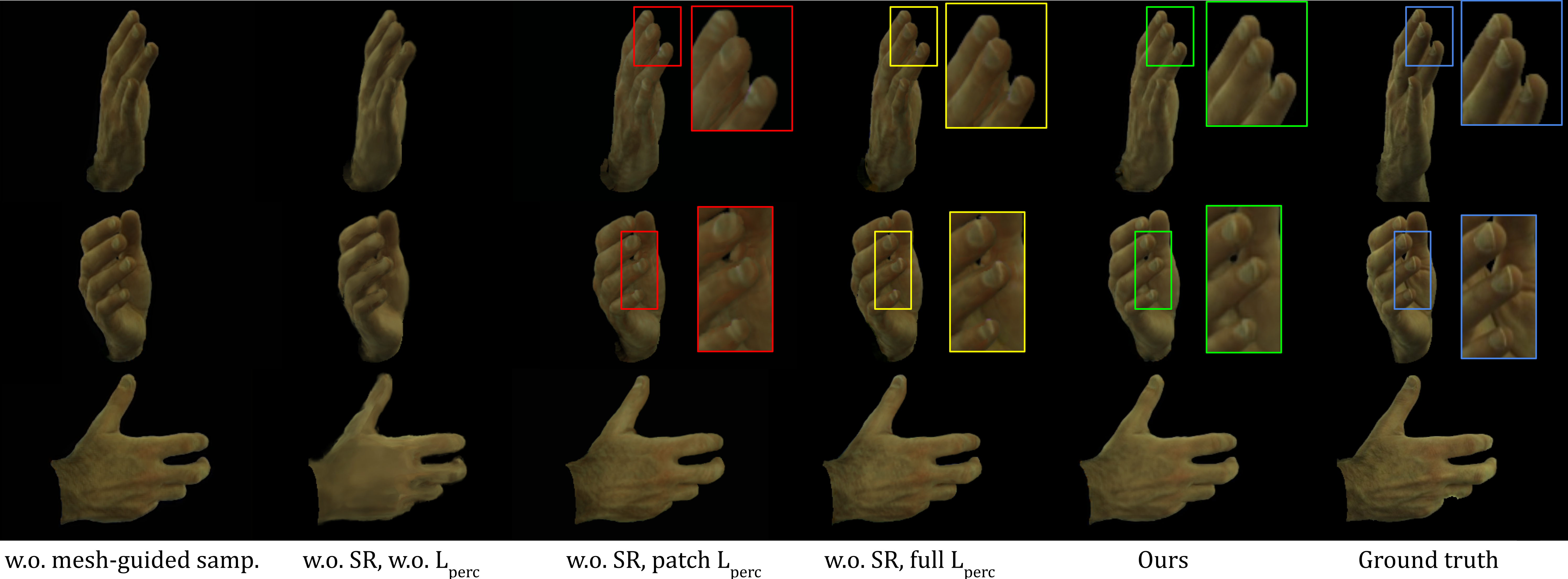}
	\caption{
 \textbf{Model Ablation.} 
 Left to right: without the mesh-guided sampling, the visual quality is good but the inference is slow (see Tab.~\ref{tab:ablation});
 without any perceptual loss, the reconstructions lack details;
 with patch-based perceptual loss, subtle artifacts appear in the details (as highlighted in red);
 with full-image perceptual loss, these details are captured correctly (as highlighted in yellow);
 finally, by using a super-resolution module, the rendering speed is further improved without compromising the details (as highlighted in green).
 }
	\label{fig:ablation}
\end{figure*}

We evaluated the impact of mesh-guided sampling by defaulting to hierarchical sampling instead (`w.o mesh-guided samp.'). 
While this produces similar rendering quality, it can be seen in Tab.~\ref{tab:ablation} that our method is $5$ times faster. 
We also evaluated the impact of the superresolution module by training our method to directly render the full-resolution image instead (`w.o. SR'). 
For this experiment, we investigated $3$ different settings:
we remove $\mathcal{L}_{perc}$ entirely (`w.o. $\mathcal{L}_{perc}$');
we implement the commonly-used patch perceptual loss ~\cite{weng_humannerf_2022_cvpr} where random crops of $64\times64$ pixels are used for the perceptual loss instead (`patch $\mathcal{L}_{perc}$'); 
finally, we use the perceptual loss on the full images (`full $\mathcal{L}_{perc}$'). 
Tab.~\ref{tab:ablation} shows that the SR module makes our method $2.4$ times faster for all variants.
Although the method `w.o. $\mathcal{L}_{perc}$' achieved the highest PSNR, adding any form of $\mathcal{L}_{perc}$ greatly increases the level of details (see Fig.~\ref{fig:ablation}). 
This increase in realism is captured quantitatively by the lower LPIPS and FID in Tab.~\ref{tab:ablation}, which better reflects human preference.
Furthermore, we show our novel application of the perceptual loss on the full image enabled by our efficient formulation (`full $\mathcal{L}_{perc}$') greatly improves the rendering quality quantitatively and qualitatively.
Finally, our full method (`Ours') achieves superior or comparable rendering quality while being significantly faster.

Overall, it is clear that our design choices optimize both rendering quality and speed, thus enabling us to photo-realistically render human hands in real-time for the first time in literature. 
Moreover, in the supplementary material, we use synthetic data to show our method's robustness to MANO fitting inaccuracies.
We also present an additional application where the hand geometry can be edited at inference time, without any additional retraining of the model.
\section{Discussion}
\subsection{Limitations and Future Work}

While our work is an important milestone for the full digitization of human hands, there are still several avenues for future work.
Since our approach depends on the MANO mesh, future work could look into improving the quality of such a mesh.
This could include refining the geometry, possibly in an end-to-end manner.
Another more strategic direction moving forward is to learn a generalizable implicit 3D morphable model of the human hands that is photoreal. 
This will give full access to all hand semantics. 
While our approach models hand-pose dependant illumination effects, it can not model shadow as a function of any random illumination condition other than the one the training set was captured under.
We leave this modeling for future works.
We hope our work encourages research into the important problem of photorealistic rendering of the human hands.

\subsection{Societal Impact}
Alongside its immense applications, human modeling also presents challenging societal problems. 
A digital avatar of an individual has the potential of being misused by bad actors. 
Though detecting real vs. fake images is a possibility, a more strategic approach would be watermarking the generative models.
This way, a generated image can always be attributed back to the model it was generated from. 
This is an active area of research, and we hope the community adopts it in their body modeling works.

\section{Conclusion} \label{sec:conclusions}

We presented the first neural implicit approach that can render human hands in a photorealistic manner in real-time. 
Our approach is carefully designed to optimize the rendering quality and speed. 
At the heart of our method is a low-resolution NeRF rendering and a super-resolution module that produces 3D-consistent results. 
We show that a novel application of the perceptual loss on the full image space is important for generating accurate details.  
We also utilize the MANO hand mesh to guide the sampling of points in 3D space to better improve the rendering speed. 
Results show that our method generates a wide variety of hand articulations, high-frequency texture details, and pose-dependent effects.
Comparison with related methods clearly shows that our approach outperforms the baselines by a significant margin.
We also demonstrate editing the hand geometry while keeping the texture fixed.
Future work could investigate learning a generalized implicit 3D morphable model of the human hands that is photoreal.

\subsection*{Acknowledgements}

We thank Ashwath Shetty, Yiming Wang, Oleksandr Sotnychenko and Basavaraj Sunagad for their help. We also thank the MPII IST department for the technical support and Jaakko Lehtinen for fruitfull discussions. This work was supported by the ERC Consolidator Grant 4DRepLy (770784).

{\small
\bibliographystyle{ieee_fullname}
\bibliography{reference}
}

\end{document}